\documentclass[conference]{IEEEtran}

\IEEEoverridecommandlockouts

\usepackage{cite}
\usepackage{amsmath,amssymb,amsfonts}
\usepackage{algorithmic}
\usepackage{multirow}
\usepackage{booktabs}
\usepackage{graphicx}

\usepackage{textcomp}
\usepackage{xcolor}
\usepackage{amsthm}  
\newtheorem{theorem}{Theorem}

\def\BibTeX{{\rm B\kern-.05em{\sc i\kern-.025em b}\kern-.08em
    T\kern-.1667em\lower.7ex\hbox{E}\kern-.125emX}}
\begin{document}

\title{Denoised Recommendation Model with Collaborative Signal Decoupling}

\author{
    \IEEEauthorblockN{Zefeng Li} 
    \IEEEauthorblockA{College of Computer Science\\
    Sichuan University\\
    Chengdu, China\\
    lizefeng0907@gmail.com} 
    
    \and 
    
    \IEEEauthorblockN{Ning Yang$^*$\thanks{ $^*$Corresponding author}} 
    \IEEEauthorblockA{College of Computer Science\\
    Sichuan University\\
    Chengdu, China\\
     yangning@scu.edu.cn} 
}

\maketitle

\begin{abstract}
Although the collaborative filtering (CF) algorithm has achieved remarkable performance in recommendation systems, it suffers from suboptimal recommendation performance due to noise in the user-item interaction matrix. Numerous noise-removal studies have improved recommendation models, but most existing approaches conduct denoising on a single graph. This may cause attenuation of collaborative signals: removing edges between two nodes can interrupt paths between other nodes, weakening path-dependent collaborative information.
To address these limitations, this study proposes a novel GNN-based CF model called DRCSD for denoising unstable interactions. DRCSD includes two core modules: a collaborative signal decoupling module (decomposes signals into distinct orders by structural characteristics) and an order-wise denoising module (performs targeted denoising on each order). Additionally, the information aggregation mechanism of traditional GNN-based CF models is modified to avoid cross-order signal interference until the final pooling operation.
Extensive experiments on three public real-world datasets show that DRCSD has superior robustness against unstable interactions and achieves statistically significant performance improvements in recommendation accuracy metrics compared to state-of-the-art baseline models.
\end{abstract}


\section{Introduction}
Nowadays, recommendation systems \cite{wu2021survey} play an increasingly important role in identifying users' genuine interests and preferences of users in different scenarios. As one of the most successful technologies in recommendation systems, collaborative filtering (CF) \cite{2007collaborative, su2009survey, 2009survey, kipf2016semi} algorithms have also been widely disseminated. Consequently, recent literature has proposed a variety of CF algorithms. In recent years, the emergence of recommendation algorithms based on Graph Neural Networks (GNNs) \cite{chen2021structured, luo2021learning} has garnered substantial attention from the research community. By stacking layers upon layers, these algorithms capture high-order signals in the user-item interaction graph. For this reason, CF algorithms based on GNNs have achieved excellent performance in recommendations.

However, due to the presence of interaction noise \cite{2006detecting, zhang2021data} in the user-item interaction graph, the user preferences we capture may not be accurate, which may further lead to a decline in the recommendation performance of the model \cite{zheng2021multi, wu2016collaborative, zhao2019collaborative}. For example, a user might mistakenly add an item to their shopping cart, or another user might accidentally like a video they are not actually interested in... These noisy interactions occur in various practical application domains of recommendation systems. In recent years, many studies have been dedicated to removing noise from the interaction graph and extracting accurate user preferences as much as possible. RGCF \cite{rgcf} integrates a graunder a multi-task learning paradigm to jointly enhance the recommendation performance, improving both robustness and recommendation diversity. GraphDA \cite{graphda} employs a pre-trained model to reconstruct a new interaction matrix and expands it by leveraging the correlations between users and between items.

Despite the efforts to remove interaction noise to some extent, existing methods do not personalize the semantic judgment of each interaction for each user. Instead, they directly delete noisy edges from the graph. This approach can weaken the collaborative signals captured by the model, leading to suboptimal recommendation performance. For example, as shown in Figure~\ref{fig1}, user $u_1$ may mistakenly interact with item $i_1$, a book. This interaction is harmful for $u_1$ and should be removed from the graph. However, if user $u_3$ is an avid reader, the collaborative signal between $u_1$ and $i_1$ could be beneficial for $u_3$. Simply deleting the edge between $u_1$ and $i_1$ would weaken the beneficial collaborative signal for $u_3$, inevitably leading to a decline in the model's recommendation accuracy.

To address the aforementioned challenges, we propose a novel GNN-based collaborative filtering method named DRCSD (Denoised Recommendation model with Collaborative Signal Decoupling). We design two primary technical modules to enable personalized semantic judgment of interactions for each node. First, unlike other GNNs that mix signals of different orders during the information propagation process, we introduce a Collaborative Signal Decoupling module to ensure that signals of different orders are propagated separately. Second, based on the first module, we design a Graph Denoising module for each order of signals to identify and remove noise signals at each level.

Our main contributions are as follows:

\begin{figure}
    \centering
    \includegraphics[width=0.5\linewidth]{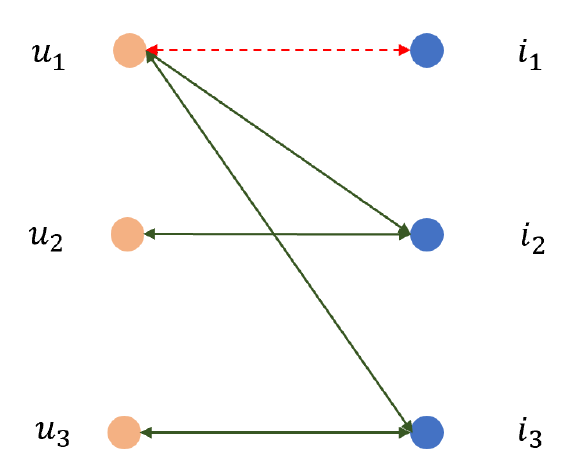}
    \caption{One possible situation: Since the interaction between $u_1$ and $i_1$ is judged as a noisy interaction, the path between $u_3$ and $u_1$ is cut off, resulting in the weakening of the cooperative signal aggregated by $u_3$. }
    \label{fig1}
\end{figure}

\begin{itemize}
\item We point out that previous denoising works have ignored the different semantics of interactions for different nodes, which leads to the sharpness of the collaborative signals captured by the model and a decrease in recommendation performance.
\item We propose a new model named DRCSD, which first decouples collaborative signals and then denoises each order of signals. This approach enables the model to remove noise from collaborative signals in a personalized manner for each node.
\item We experiment with our proposed DRCSD of six backbones on three popular datasets. And extensive experimental results demonstrate the effectiveness and generalization of our framework.

\end{itemize}
Due to space constraints, related work and preliminary are provided in the Appendix.

\begin{figure*}[t]
    \centering
    \includegraphics[width=1\linewidth]{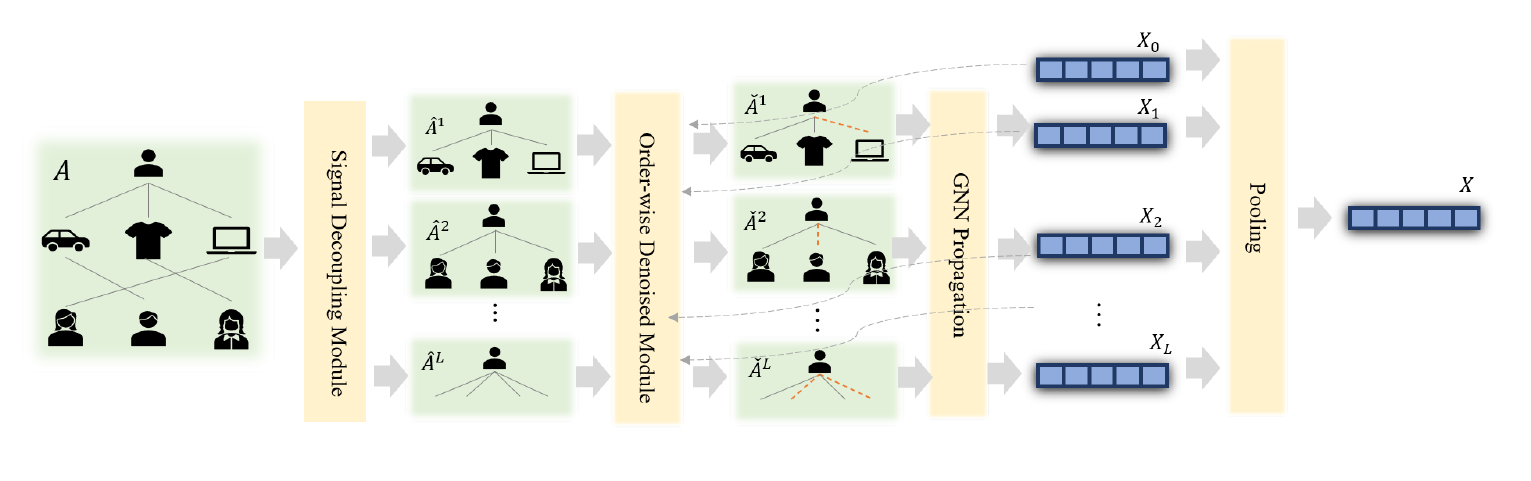}
    \caption{Overview of DRCSD model, except classical GNN module,  DRCSD model also contains the signal decoupling module, the order-wise denoised module. Signal decoupling module decouples each order collaborative into different interaction grpah, and oder-wise denoised module removes the noise in the signals of each order. In addition, we modify the classical GNN propagation mechanism to prevent the mixing of signals of all orders during propagation.  }
    \label{fig2}
\end{figure*}
\section{Methodology}
In this section, we will introduce our proposed model DRCSD. 

\subsection{Overview}
To achieve personalized noise removal for each node, we designed two modules: the signal decoupling module and the order-wise denoising module. In the signal decoupling module, we decouple the initial interaction matrix A to ensure that the decoupled $l$-th order interaction matrix contains only the $l$-th order collaborative signals \cite{rabiah2024bridging,sun2024collaborative, he2024layer}. In the order-wise denoising module, we first use the node representations obtained from the convolution of the first $l-1$ orders to compute the hidden layer states $H_l$ of the $l$-th order. We then calculate the similarity between interacting nodes using the hidden layer states which serves as the basis for identifying noisy edges. Finally, we employ the Gumbel-softmax method to sample and generate a masking matrix $W$.
It is worth noting that we have modified the message propagation mechanism of the GNN-based model to prevent the mixing of collaborative signals of different orders during propagation \cite{yang2020graph}.
\subsection{Signal Decoupling Module}
Since one single graph convolution operation \cite{chen2024improving} in Graph Neural Networks (GNNs) can only aggregate signals from the first-order neighborhood, multiple graph convolution operations are required to aggregate higher-order signals. Therefore, in the Signal Decoupling Module, we first perform signal decoupling on the initial interaction graph $A$ to ensure that the decoupled $l$-th order interaction matrix $A_l$ contains only the $l$-th order collaborative signals. 

Here, we define that the $l$-th order collaborative signals referred to in this paper represent the signals from neighbor nodes that are $l$ hops away from the target node.

\begin{theorem}
Let $G=(V,E)$ be a graph, where $V$ is the set of vertices and $E$ is the set of edges. For any two vertices $u,v \in V$, define $dG(u,v)$ as the length of the shortest path between $u$ and $v$ in $G$. For any positive integer $l$, define $G^l$ as the $l$-th power of $G$, in which two vertices $u$ and $v$ are adjacent in $G^l$ if and only if 
\begin{equation}
    dG(u,v) \leq l
\end{equation}
\end{theorem}
Therefore, we decouple the interaction matrix of the l-th order $\hat{A}^l$ is as follows:
\begin{equation}
    \hat{a}^l_{(u,i)}=
    \begin{cases}
        k, & \text{if }\hat{a}^{t}_{(u,i)} =0 \\
        0, & \text{else}
    \end{cases}
\end{equation}
where  $\hat{a}^l_{(u,i)}$ is the value between node $u$ and $v$ in $\hat{A}^l$ and k refers to the number of the ways that connect $u$ and $v$, and $t = l-1, l-2, ...,1$.

According to theorem 3.1, we can know that $\hat{a}^l_{(u,i)} >0$ means there are paths of length less than or equal to l between node u and node v. While $\hat{a}^{l-1}_{(u,i)} =0$, we can make sure that the length of the paths between $u$ and $v$ is greater than $l$. In that case, we can promise the shortest length of paths between $u$ and $v$ equals $l$.

\subsection{Order-wise Denoised Module}
After we obtain the interaction matrix of each order, what we need to do is to remove the noise of each order. In this section we will describe how we eliminate noise in each order.

\begin{equation}
    H_l=P(X_0, X_1, ...,X_{L-1})
\end{equation}

Since the node representations of the previous $l - 1$ orders describe the different states of nodes under different domains, we regard the embedding expression obtained by fusing the embeddings from the previous convolution as the hidden representation of the $l$-th order. 

\begin{equation}
    cos(h_u^l,h_v^l) = \frac{h_u^Th_v}{\left \| h_u \right \|_2
    \left \| h_v \right \|_2}
\end{equation}

\begin{equation}
    s_{u,v}=\frac{ cos(h_u^l,h_v^l)+1}{2}
\end{equation}

Next, we calculate the cosine similarity \cite{xia2015learning} between connected nodes using the hidden states of the nodes. To normalize the similarity to the interval [0,1], formula (8) applies a linear transformation to the cosine similarity, resulting in the similarity score $s_{u,v}$. Likewise, we define the corresponding dissimilarity as $1-s_{u,v}$.

The intrinsic basis for judging noise is that in an interaction between two nodes, the more similar the hidden states of the two nodes are, the higher the reliability of the interaction, and the lower the likelihood that the interaction is noise.

\begin{equation}
    p = [s_{u,v}, dis_{u,v}]
\end{equation}
    
\begin{equation}
   w^l_{u,v} = Gumbel-softmax(p,\gamma)
\end{equation}

We introduced the Gumbel-Softmax \cite{jang2016categorical, herrmann2020channel} method, which is primarily used to sample from a vector p that contains similarity and dissimilarity information. This method allows us to generate a weight $w_{u,v}^l$ for each edge that reflects its degree of noise.

\begin{equation}
   \check{A}^l = W^l\odot\hat{A}^l
\end{equation}

Finally, we use the mask matrix $W^l$ and the interaction matrix of the l-th order to perform the Hadamard product \cite{million2007hadamard}, resulting in the denoised interaction matrix of the l-th order.

\subsection{New Message Propagation}
Although we have decoupled the initial adjacency matrix and ensured that it contains only signals of specific orders, traditional Graph Neural Networks (GNNs) mix signals \cite{zhang2018hybrid} of different orders during the message passing process. This mixing of signals hinders our goal of denoising signals of different orders separately. Therefore, it is necessary to improve the existing GNN message passing mechanism and propose a new mechanism.

\begin{equation}
    X_l = \tilde{A}^lX_0
\end{equation}
where $\tilde{A}^l=\check{D}^{-\frac{1}{2}}\check{A}\check{D}^{-\frac{1}{2}}$  is the denoised interaction matrix of the l-th order after symmetric normalization

Specifically, we modify Equation (2) to Equation (12). For the convolution formula of the l-th order, we directly use the initialized embedding matrix $X_0$ to perform the graph convolution operation. This ensures that the resulting embedding matrix $X_l$ contains only the collaborative signals of the l-th order.
\begin{equation}
    X =P(X_0,X_1,...,X_L )
\end{equation}
Finally, we use the pooling function \cite{lee2016generalizing}to aggregate the whole orders embedding matrix and obtain the final representation $X$.

\subsection{Model Training}
 We train each module via a multi-task learning strategy\cite{sener2018multi} . First, we use the BPR loss to train the recommendation.

\begin{equation}
    L_{BPR} = -log\sigma(x_ux_{i^+} - x_ux_{i^-})
\end{equation}

where $\sigma()$ denotes the sigmoid function, $x_u$ refers to embedding of the user u. $x_{i^+}$ denotes to the embedding of the item that user u actually interact with, and $x_{i^-}$ denotes the embedding of the item that u does not interact with.

\begin{equation}
    L_{d} =\sum_{l=1}^{l} \|\tilde{A}^l- \overline{A}^l\|_1 
\end{equation}

Then, we design the $L_d$ loss to limit the difference between the denoised graph and the denoised graph to not be too large. where $\overline{A}^l=\hat{D}^{-\frac{1}{2}}\hat{A}\hat{D}^{-\frac{1}{2}}$,

\begin{equation}
    L = L_{BPR} + \beta L_d + \gamma \|\Theta\|_2^2
\end{equation}

where denoised cofficient $\beta$ and $\gamma$ are hyperparameters to control the strength of the $L_d$ loss and $L_2$ regularization respectively, and $\Theta= \{X_U , X_I\}$ is the set of model parameters.

To improve the performance of our model, we combine the BPR loss, $L_d$ loss and $L_2$ regularization loss \cite{cortes2012l2} to obtain the final loss $L$, and use the $L$ loss for end-to-end training. The analysis of the time and space complexity is provided in the Appendix C.

\section{Experiment}
In this section, we will demonstrate the effectiveness and robustness of our proposed model by different kinds of experiments.

\begin{table}[h]
\centering
\caption{Dataset Statistics }
\label{tab1}
\begin{tabular}{ccccc}
\toprule
Dataset & \#Users & \#Items & \#Interactions & Sparsity \\
\midrule
Retailrocket\_ADDtocart & 37,722 & 23,903 & 62,025 & 99.99\% \\
Retailrocket\_Transaction & 11,719 & 12,025 & 21,270 & 99.98\% \\
Lastfm & 1,892 & 17,632 & 92,834 & 99.72\% \\
\bottomrule
\end{tabular}
\end{table}

\subsection{Experimental Setup}

\subsubsection{Dataset} We use three real-world recommendation datasets (Retailrocket\_ADDtocart, Retailrocket\_Transaction and Lastfm) for evaluation.

To simulate real-world implicit noise, we treat all dataset interaction records as positive feedback and summarize their statistics in Table~\ref{tab1}. Each dataset’s interactions are randomly split into training/validation/test sets at a 7:1:2 ratio, with one negative example sampled per positive example during training.

For simulating malicious noise injection, we artificially added 5\%, 10\%, 15\%, and 20\% noise (non-existent interactions) to three real datasets to generate synthetic datasets. To ensure reliability, noise was only injected into the training set, with the validation and test sets unchanged.



\subsubsection{Baseline Models} We compare the performance of the proposed model with the following baseline models: BPRMF\cite{mf}, NGCF\cite{ngcf}, DGCF\cite{dgcf}, LightGCN\cite{lightgcn}, SGL\cite{sgl}, HMLET\cite{hmlet}, and RGCF\cite{rgcf}. Detailed descriptions of the baseline models are provided in Appendix D.

\subsubsection{Implementation Details}Two implementations were adopted for all baseline models: reproduction via authors’ source code and utilization of the Recbole library. Hyperparameters were configured with node embedding dimension = 64, batch size = 2048, and remaining parameters adjusted per original papers.
All models were trained with the Adam optimizer. To mitigate overfitting, a maximum patience of 10 was set—training terminates early if validation performance plateaus for 10 consecutive epochs. Model performance was evaluated using Recall@20, Precision@20, and NDCG@20, with average results across all users reported for reliability.


In this study, for the proposed model, we have meticulously set the key parameters. Through theoretical analysis and preliminary experiments, the range of the number of model layers has been determined to be {2, 3}. The range of the learning rate is set to [1e-5, 1e-3] to ensure convergence efficiency and accuracy. In addition, verified by multiple sets of experiments, the range of the parameter beta is {0.3, 0.4, 0.5}, and the range of the parameter gamma is {1e-3, 1e-4, 1e-5}, which provides a reasonable configuration for the optimization of the model.

\begin{table*}[!t]
    \caption{The overall performance comparison on three real-world datasets and three synthesized datasets. The best result is bolded and the runner-up is underlined. $^\ast$ indicates the statistical significance for $p < 0.05$ compared to the best baseline.}\label{tab2}

    \vspace{-5mm}
    \begin{center}
        \renewcommand{\arraystretch}{1}
        \tabcolsep=0.12cm
        \resizebox{\textwidth}{!}
{
\begin{tabular}{clccccccccc}
\hline
Dataset & Metric & BRRMF & NGCF & DGCF & LightGCN & SGL & HMLET & RGCF & DRCSD & Improv. \\
\hline
\multirow{3}{*}{Retailrocket\_ADDtocart} 
& Recall@20 & 0.0377 & 0.0627 & 0.0623 & 0.0630 & \underline{0.0709} & 0.0641 & 0.0644 & \textbf{0.0733 $^\ast$} & +3.27\%\\
& NDCG@20 & 0.0203 & 0.0389 & 0.0387 & 0.0400 & \underline{0.0420} & 0.0407 & 0.0378 & \textbf{0.0434 $^\ast$}& +3.23\%\\
& Precision@20 &0.0022 & 0.0048 & 0.0049 & 0.0050 & \underline{0.0054} & 0.0051 & 0.0050 & \textbf{0.0058 $^\ast$} & +6.89\% \\
\hline
\multirow{3}{*}{Retailrocket\_Transaction} 
& Recall@20 & 0.0345 & 0.0409 & 0.0426 & 0.0446 & 0.0444 & \underline{0.0459} & 0.0450 & \textbf{0.0467 $^\ast$} & +2.56\%\\
& NDCG@20 & 0.0227 & 0.0309 & 0.0310 & 0.0331 & \underline{0.0345} & 0.0341 & 0.0336& \textbf{0.0355 $^\ast$} & +2.89\%\\
& Precision@20 & 0.0023 & 0.0029 & 0.0028 & 0.0030 & 0.0031 & 0.0032 & \underline{0.0034} & \textbf{0.0035 $^\ast$} & +2.86\% \\
\hline
\multirow{3}{*}{Lastfm} 
& Recall@20 & 0.2267 & 0.2449 & 0.2480 & 0.2583 & 0.2599 & 0.2599 & \underline{0.2609} & \textbf{0.2623 $^\ast$} & +0.53\% \\
& NDCG@20 & 0.2129 & 0.2327 & 0.2367 & 0.2460 & 0.2488 & 0.2490 & \underline{0.2496} & \textbf{0.2508 $^\ast$} & +0.47\%\\
&Precision@20 & 0.1110 & 0.1203 & 0.1222 & 0.1270 & 0.1283 & 0.1279 & \underline{0.1288} & \textbf{0.1297 $^\ast$} & +0.70\%\\
\hline
\end{tabular}
}
    \end{center}
    \label{tab:attack_result}
    \vspace{-3mm}
\end{table*}
\subsection{OverPerformance}
Experimental results of DRCSD and baseline models are summarized in Table~\ref{tab2}, where DRCSD achieves the best performance across all datasets. Further analysis shows GNN-based models outperform MF-based counterparts, especially on the first two sparser datasets—demonstrating GNNs’ advantage in leveraging graph structure for effective node embedding learning in sparse data. Robust recommendation models generally outperform GNN-based ones, indicating dataset interaction noise and robust models’ noise-filtering effectiveness. Most notably, DRCSD comprehensively outperforms robust models, as its node-specific personalized noise removal design eliminates noise interference more effectively.

\begin{figure}
    \centering
    \includegraphics[width=0.95\linewidth]{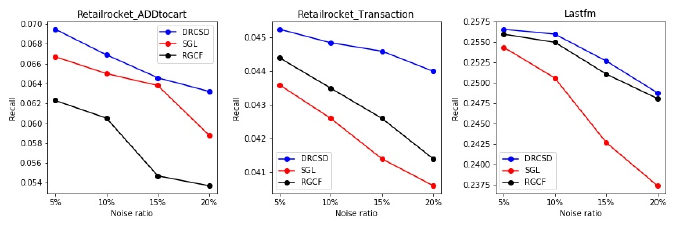}
    \caption{The running results of the DRCSD, SGL and RGCF models on three datasets with gradually increasing noise from 5\% to 20\% }
    \label{fig3}
\end{figure}

\begin{figure}[!t]
    \centering
    \includegraphics[width=0.95\linewidth]{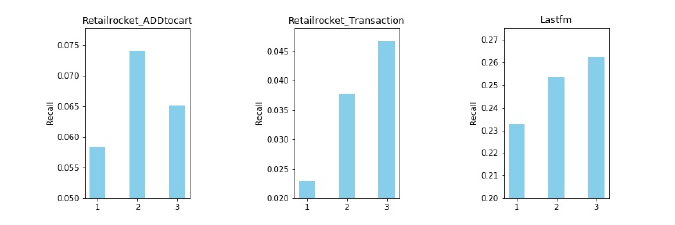}
    \caption{DRCSD performance of different layers L }
    \label{fig4}
\end{figure}

Synthetic datasets are constructed by incrementally injecting interaction noise (5\%–20\%) into three real-world datasets. Figure~\ref{fig3} presents the performance of SGL, RGCF, and DRCSD under varying noise ratios—all models show gradual performance degradation with increasing noise. Notably, DRCSD consistently outperforms the two robust baselines on noisy datasets, demonstrating its superior noise resistance.

\subsection{Further Analysis}
To further validate the robustness and efficacy of DRCSD, we conduct a series of in-depth analyses.
\subsubsection{Hyperparameter study}
To systematically investigate the influence of distinct hyperparameter values on model performance, the performance metrics of the model under varying hyperparameter configurations are tabulated in Figure~\ref{fig3} and Figure~\ref{fig4}.

\begin{figure}[htbp]
    \centering
    \includegraphics[width=0.95\linewidth]{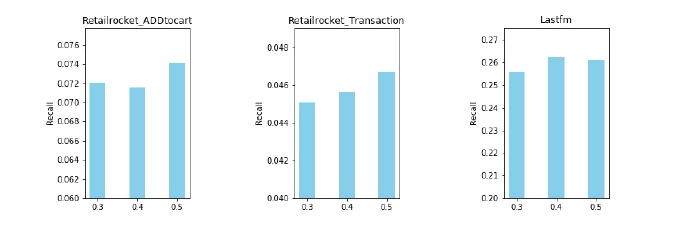}
    \caption{DRCSD performance of different denoised coefficient $\beta$}
    \label{fig5}
\end{figure}
Figure~\ref{fig4} investigates the impact of layer configurations on model performance. Initial increases in layers yield significant performance improvements, underscoring enhanced information aggregation for deeper feature learning and comprehensive graph topology capture. However, beyond a threshold, performance degrades on specific datasets—attributed to over-smoothing induced by reduced node-wise dissimilarity from excessive depth.

Figure~\ref{fig5} explores the denoising coefficient $\beta$ (a key hyperparameter regulating denoising intensity). Results show optimal $\beta$ varies by dataset, corroborating distinct implicit noise proportions across datasets. Thus, dataset-specific selection is imperative for optimal performance.


\begin{table}[htbp]
\centering
\caption{Ablation results on three datasets}\label{tab3}
\begin{tabular}{clcccc}
\hline
\scriptsize Dataset & \scriptsize Metric & \small w/o DM & \small w/o SD & \small DRCSD \\
\hline
\scriptsize \multirow{3}{*}{Retailrocket\_ADDtocart} 
& \scriptsize Recall@20 & \small 0.0664 & \small 0.0698 & \small \textbf{0.0741} \\
& \scriptsize NDCG@20 & \small 0.0423 & \small 0.0417 & \small \textbf{0.0451} \\
& \scriptsize Precision@20 & \small 0.0051 & \small 0.0054 & \small \textbf{0.0058} \\
\hline
\scriptsize \multirow{3}{*}{Retailrocket\_Transaction} 
& \scriptsize Recall@20 & \small 0.0441 & \small 0.0449 & \small \textbf{0.0467} \\
& \scriptsize NDCG@20 & \small 0.0343 & \small 0.0353 & \small \textbf{0.0355} \\
& \scriptsize Precision@20 & \small 0.0032 & \small 0.0033 & \small \textbf{0.0034} \\
\hline
\scriptsize \multirow{3}{*}{Lastfm} 
& \scriptsize Recall@20 & \small 0.2547 & \small 0.2550 & \small \textbf{0.2623} \\
& \scriptsize NDCG@20 & \small 0.2453 & \small 0.2440 & \small \textbf{0.2508} \\
& \scriptsize Precision@20 & \small 0.1257 & \small 0.1261 & \small \textbf{0.1297} \\
\hline
\end{tabular}
\end{table}

\subsubsection{Ablation study}
Ablation experiments were conducted to validate each DRCSD module’s effectiveness: "w/o DM" removes denoising modules from all layers, while "w/o SD" omits signal decoupling (resulting in multi-order collaborative signal mixing during propagation). As shown in Table~\ref{tab3}, removing denoising modules degrades performance across all datasets—confirming dataset interaction noise, the necessity of denoising, and the modules’ effectiveness in noise identification and elimination. Additionally, performance degradation without signal decoupling verifies that the proposed node-personalized denoising outperforms conventional direct interaction noise removal.

\section{Conclusion}
This paper proposes DRCSD, a novel denoising recommendation model comprising two core components: Signal Decoupling Module and Order-wise Module. We optimize traditional GNN forward propagation to avoid mixing multi-order collaborative signals— the Signal Decoupling Module decouples signals to ensure graph convolution aggregates only specific-order signals, which are then fed into the Order-wise Module for node-personalized denoising. A self-supervised task is designed to constrain noise tolerance and aid model optimization. Extensive experiments on three real-world and multiple noisy synthetic datasets show DRCSD outperforms all baselines across metrics, verifying its superior noise resistance.
Future work will extend this study, e.g., enhancing the model’s data modeling capability to handle complex graph structures and larger datasets.

\section{Acknowledgements}
This work is supported by Natural Science Foundation of Sichuan Province under grant 2024NSFSC0516 and National Natural Science Foundation of China under grant 61972270.
\bibliographystyle{IEEEtran}
\bibliography{references}

\clearpage
\appendices
\section{Related work}
In this section, we will provide a concise overview of the relevant literature in the area of graph collaborative filter and denoised recommendation.
\subsection{Graph collaborative filter}
As the core technology of modern recommendation, Collaborative Filtering (CF) is dedicated to mining user preferences from the historical interaction data between users and items, thereby achieving precise personalized recommendation services \cite{chen2025research}. In recent years, Graph-based Collaborative Filtering (GCF) technology has gradually become a research hotspot and mainstream paradigm in this field. Its essence lies in leveraging the graph structure formed by user-item interaction data to deeply analyze the complex relationships between users and items, thus significantly enhancing the effectiveness and accuracy of recommendation models \cite{lightgcn, ngcf}.

In the early development stage of collaborative filtering technology, methods such as Matrix Factorization (MF) \cite{mf} were widely used. These methods map users and items to a shared low-dimensional latent space and predict the degree of user preference for items through vector inner product operations. However, these traditional methods have significant limitations. On the one hand, they struggle to fully exploit the high-order relationships and complex structural information contained in user-item interaction data. On the other hand, in scenarios with high data sparsity, the prediction accuracy of the models is severely affected. With the increasing maturity of Graph Neural Networks (GNNs) technology, it has brought new development opportunities to the field of collaborative filtering. The Neural Graph Collaborative Filtering (NGCF) \cite{ngcf} model innovatively propagates collaborative information on the user-item bipartite graph, effectively modeling the high-order connectivity between users and items, and deeply integrating collaborative signals into the embedding process, which greatly enhances the expressive ability of user and item representations. The subsequently proposed LightGCN \cite{lightgcn} model further optimizes and simplifies NGCF by retaining only the neighborhood aggregation mechanism in graph convolution operations. While reducing the model complexity, it significantly improves the model's applicability and achieves excellent recommendation performance in numerous practical application scenarios.

Despite the significant progress made in GNN-based graph collaborative filtering technology, it still faces many key issues that urgently need to be addressed. Currently, mainstream GNN-based GCF methods generally adopt multi-layer GNN architectures to achieve the propagation and aggregation of high-order neighbor information. However, this approach not only leads to a substantial decrease in computational efficiency, making it difficult to meet the requirements of large-scale real-time recommendation scenarios, but also may introduce model bias due to the inclusion of irrelevant neighbor information. In addition, in practical applications, due to the highly sparse nature of user-item interaction data, the widely used Bayesian Personalized Ranking (BPR) loss function often fails to provide sufficient supervision signals, thus affecting the model training effect. Meanwhile, noise data often exists in user-item interaction data, such as user misoperations, malicious ratings, and errors during data collection. These noises interfere with the model's learning of real user preferences and item features, making it difficult for the model to accurately identify valid information. Existing graph collaborative filtering algorithms lack efficient and precise solutions for data denoising and are unable to effectively purify data in complex data environments, thereby affecting the accuracy and reliability of recommendations.

\subsection{Denoised Recommendation}

In the field of recommendation systems, user-item interaction data constitutes the core foundation of personalized recommendations. However, in practical applications, such data is often subject to various types of noise interference, such as user mis-clicks, malicious ratings, and errors in data collection. These noise factors make it difficult for recommendation models to accurately capture users' true preferences and the intrinsic features of items, thereby significantly affecting the accuracy and reliability of recommendation results. Denoising recommendation technology has emerged to address this core issue.

In recent years, research on denoising recommendation technology has made significant progress, giving rise to numerous innovative models. Among them, the Robust Graph Collaborative Filtering (RGCF)\cite{rgcf} model and the Graph Collaborative Signal Denoising Enhancement (GraphDA) model are particularly noteworthy. The RGCF model, based on graph structure, calculates the reliability scores of interactions through a graph denoising module. It then performs hard denoising (i.e., deletion) on low-reliability interactions and soft denoising weighting on the remaining interactions. Meanwhile, the model utilizes a diversity preservation module to construct an enhanced graph, balancing the accuracy and diversity of recommendations. The GraphDA model \cite{graphda} focuses on addressing data noise and sparsity issues. It first conducts pre-training embedding with LightGCN, then generates a denoised interaction matrix through Top-K sampling, and introduces user-user and item-item correlation matrices to mitigate the diffusion of noise in higher-order message propagation.

Despite the achievements of these models, current denoising recommendation systems still face significant challenges. Most systems adopt a uniform denoising strategy for all nodes in the interaction graph, without fully considering the differences in interaction patterns and data distribution among different nodes. This "one-size-fits-all" denoising approach is highly likely to suppress effective collaborative signals during the denoising process. For example, over-denoising sparse but critical interactions can lead to the loss of important preference information, while under-denoising noisy nodes fails to correct misleading signals. Ultimately, these issues weaken the model's ability to accurately model user preferences and item features, resulting in a decline in the performance of recommendation systems. Therefore, achieving personalized noise removal for each node is a crucial key to enhancing the effectiveness of denoising recommendations.

\section{Preliminary}

In this section, we will provide an introduction to GNN-based recommendation system models and denoising recommendation models, which will help us gain a better understanding of the model we propose. 

\subsection{GNN-based Recommendation Model}

With the rise of deep learning, deep learning based \cite{wei2017collaborative, lee2022deep, kipf2016semi},  methods have seen a rapid increase in their applications within the field of recommendation systems. Graph Neural Networks (GNNs) \cite{wu2020comprehensive}  have garnered significant attention due to their ability to extract accurate node representations from graph data \cite{li2020hierarchical}. Given that the interaction information between users \textit{U} and items \textit{I} can be represented as graph data, GNNs are capable of capturing high-order signals \cite{giraldo2024higher, jin2022graph} in collaborative filtering algorithms.

In GNN-based recommendation system models, user-item interactions can be represented as a bipartite graph \textit{A = \{V, E}\} . In this bipartite graph, each user and item is regarded as a node, while their interactions are represented as edges connecting the corresponding nodes.
\begin{equation}
A=\begin{bmatrix} 0 & R \\ R & 0 \end{bmatrix}
\end{equation}
where $R \in \{0,1\}^{|U|*|I|} $  is the user-item interaction matrix \cite{kunal2023gnn}.

When GNN works, its basic idea is aggregating the features \cite{gama2019aggregation, parada2023stability} of the neighborhood nodes on the graph and represent them as new nodes. Its first-order aggreagation formulation is expressed as follows:

\begin{equation}
X^l=\tilde{A}X^{l-1}
\end{equation}
where $\tilde{A} = D^{-\frac{1}{2}}AD^{-\frac{1}{2}}$, $X^l$ representing the node representation for l-th layer, and $D$ is the degree matrix of $A$.Since a single graph convolution operation can only aggregate first-order collaborative signals, it is necessary to iteratively apply Equation (2) multiple times to capture higher-order collaborative signals and obtain node representations $X^L$ for L-th layer.And then We input the embedding expressions \cite{liu2020gnn, cai2018comprehensive}of nodes at each layer into the pooling function to obtain the final embedding expressions.
\begin{equation}
   X = Pool(X^0,X^1, X^2,...,X^L)
\end{equation}

Finally, we measure the possibility of user $u$ interacting with item $i$ by computing the inner product of their final representations.

\begin{equation}
    y_{u,i}=x_u^Tx_i
\end{equation}

where $x_u$ and $x_i$ is the final representation of node user $u$ and item $i$, which are obtained by fusion \cite{li2019graph2seq, yang2024gl} of representations of each layer.

\subsection{Noise in GNN-Based Recommendation}
In the last subsection, we pointed out that GNN-based recommendation systems generate the representation of each node by aggregating neighborhood information in the interaction graph. As can be seen from Equations (1) and (2), the quality of the interaction graph A is directly related to the quality of the node representations generated by the model.

However, in the real world, noise in user-item interaction graphs is a common phenomenon \cite{ping2024ddrec,sun2024self}. In this paper, we define noise as interactions that do not reflect users' true preferences. These interactions may arise accidentally due to users' erroneous clicks \cite{hu2024hierarchical, wang2024guiding}, or they may be the result of malicious attackers \cite{ma2024stealthy} injecting false interaction information into the records. Numerous studies have pointed out that noisy information is detrimental to the model's ability to capture genuine collaborative signals and will reduce the recommendation accuracy of recommendation systems.

\section{Time and Space Complexity} 
The primary time complexity of the DRCSD model originates from two components: the computation of noise at each layer and the calculation of losses. It is worth noting that the operation of decoupling signals of various orders in the signal decoupling module is precomputed before the start of model training and therefore should not be included in the time complexity calculation. For the computation of noise at each layer, the time complexity per training epoch is $O(dL\lvert\mathcal{E}\rvert)$. In terms of loss calculation, the time complexity for computing the BPR loss is $O(d\lvert\mathcal{E}\rvert+dZN)$, where $z$ represents the number of negative example pairs sampled for each positive example pair during BPR loss computation, and $z$ is typically much smaller than $N$. The time complexity required for computing the $L_d$ loss is  $O(d\lvert\mathcal{E}\rvert)$, while the time complexity for computing the $L_2$ regularization loss is $O(dN)$. Consequently, the total time complexity of the DRCSD model is $O(dL\lvert\mathcal{E}\rvert+d\lvert\mathcal{E}\rvert+dN +dZN)$. In contrast, the time complexity required for each training epoch of the LightGCN\cite{lightgcn} model is $O(dL\lvert\mathcal{E}\rvert+d\lvert\mathcal{E}\rvert)$. Since the time complexities of the DRCSD model and LightGCN model remain at the same order of magnitude, this demonstrates that the proposed DRCSD model is an efficient and effective recommendation system model.

\section{Baseline Models}

1. \textbf{MF based model}:
\begin{enumerate}
    \item BPRMF \cite{mf} is a matrix factorization recommendation model based on the idea of Bayesian personalized ranking.
    
\end{enumerate}

2. \textbf{GNN based model}:
\begin{enumerate}
    \item NGCF \cite{ngcf} learns the embedding representations of users and items by aggregating the neighbor node information of users and items, thereby achieving more accurate personalized recommendations action of users and items. So as to achieve more accurate personalized recommendations.
    \item DGCF \cite{dgcf} improves the performance of collaborative filtering by decoupling user intentions and decouples the representations of users and items at the granularity of user intentions, thereby generating more accurate personalized recommendations.
    \item LightGCN \cite{lightgcn} is an efficient and concise graph neural network recommendation model that does not contain any learnable weight parameters, thereby achieving fast and accurate personalized recommendations.
\end{enumerate}

3.  \textbf{Denoised model}:
\begin{enumerate}
     \item SGL \cite{sgl} is a recommendation system model based on contrastive learning, which enhances the representation learning of graph-structured data in a self-supervised manner.
    \item HMLET \cite{hmlet} optimizes the embedding representation of users and items by dynamically selecting the linear or nonlinear propagation mode through the gating module, thereby improving the accuracy and robustness of the recommendation system.
    \item RGCF \cite{rgcf} reduces the influence of noise interaction through the graph denoising module and enhances the recommendation diversity by using the diversity preservation module, thereby improving the robustness and accuracy of the recommendation system.
\end{enumerate}

\end{document}